\documentclass[sigconf]{acmart}

\acmConference[MobiHoc'18]{the Nineteenth International Symposium on Mobile Ad Hoc Networking and Computing}{June 2018}{Los Angeles, California USA}
\acmYear{2018}
\copyrightyear{2018}

\usepackage[noend]{algorithmic}
\usepackage{algorithm}
\usepackage{amsfonts}
\usepackage{amsmath}
\usepackage{amstext}
\usepackage{balance}
\usepackage{gensymb}
\usepackage{graphicx}
\usepackage{pgfplots}
\usepackage{subfigure}
\usepackage{times} 
\usepackage{xspace}
\pgfplotsset{compat=1.4}
\usepackage[normalem]{ulem}

\newtheorem{thm}{Theorem}
\newtheorem{defn}{Definition}
\newtheorem{cor}[thm]{Corollary}

\newenvironment{myproof}{\noindent {\bf Proof:  }}{\hfill\rule{2mm}{2mm} \vskip \belowdisplayskip}
\newenvironment{proofof}[1]{\noindent{\bf Proof of #1:  }}{\hfill\rule{2mm}{2mm}\vskip \belowdisplayskip}

\newcommand{\extremept}{extreme point\xspace}
\newcommand{\extremepts}{extreme points\xspace}

\newcommand{\ie}{i.e.,\xspace}
\newcommand{\eg}{e.g.,\xspace}
\newcommand{\aka}{aka\xspace}
\newcommand{\wrt}{w.r.t.\xspace}
\newcommand{\etal}{et al.\xspace}
\newcommand{\idp}{IDP\xspace}

\newcommand{\ints}{\mathbb{Z}}
\newcommand{\R}{\mathbb{R}}
\newcommand{\Rm}{\R^m} 

\newcommand{\dimm}{m}
\newcommand{\Span}[1]{\mathrm{span} (#1)}
\newcommand{\half}{\frac{1}{2}}
\newcommand{\graph}{G}
\newcommand{\nodes}{V}
\newcommand{\edges}{E}
\newcommand{\opt}{OPT}
\newcommand{\dB}{\mathrm{dB}}

\newcommand{\src}{s}
\newcommand{\dest}{t}
\newcommand{\st}{$\src$-$\dest$\xspace}

\newcommand{\Path}{P}
\newcommand{\loss}{\ell}
\newcommand{\dist}{d}
\newcommand{\pair}{(\dist, \loss)}
\newcommand{\pointcloud}{\mathcal{P}}
\newcommand{\logmult}{\alpha}
\newcommand{\PL}{\mathrm{PL}}
\newcommand{\dpobj}{f} 
\newcommand{\wallloss}{\mathrm{WL}}
\newcommand{\diffracloss}{\mathrm{DL}}
\newcommand{\Angle}{\theta}
\newcommand{\diffconst}{\delta}
\newcommand{\fsexp}{\gamma}
\newcommand{\obj}{f}
\newcommand{\levelcurve}{\mathcal{F}}

\newcommand{\param}{\lambda}

\newcommand{\SP}{\mathrm{SP}} 
\newcommand{\lossp}{\loss_{\param}}
\newcommand{\distp}{\dist_{\param}}
\newcommand{\pairp}{(\distp, \lossp)}
\newcommand{\gOne}{G_1}
\newcommand{\nodesOne}{\nodes_1}
\newcommand{\edgesOne}{\edges_1}
\newcommand{\gTwo}{G_2}

\newcommand{\edgesTwo}{\edges_2}
\newcommand{\corners}{C}
\newcommand{\corner}{c}
\newcommand{\dests}{T}
\newcommand{\numBreaks}{k}
\newcommand{\dijkstralabel}{L}
\newcommand{\weight}{w}

\newcommand{\gpratio}{r}

\newcommand{\GP}[2]{GP(#1, #2)}
\newcommand{\GPd}{\GP{\gpratio}{\param_0}}
\newcommand{\unif}{u}
\newcommand{\activeinterval}{I}
\newcommand{\mpts}{M}

\newcommand{\optP}{\Path^*}
\newcommand{\optdist}{\dist^*}
\newcommand{\optloss}{\loss^*}
\newcommand{\optpair}{(\optdist, \optloss)}
\newcommand{\optparam}{\param^*}
\newcommand{\paramlo}{\param_{\mathrm{lo}}}
\newcommand{\paramhi}{\param_{\mathrm{hi}}}
\newcommand{\zeroinfty}{[0, \infty]}
\newcommand{\Hybrid}[1]{h_{#1}}
\newcommand{\Hparam}{\Hybrid{\param}}
\newcommand{\Pparam}{\Path_{\param}}
\newcommand{\dparam}{\dist_{\param}}
\newcommand{\lparam}{\loss_{\param}}
\newcommand{\dlparam}{(\dparam, \lparam)}
\newcommand{\dmin}{\dist_{\min}}
\newcommand{\dmax}{\dist_{\max}}
\newcommand{\Dmin}{D_{\min}}
\newcommand{\Dmax}{D_{\max}}
\newcommand{\lineptslope}[3]{\mathcal{L}(#1, #2, #3)}
\newcommand{\opttangent}{\lineptslope{\optdist}{\optloss}{\optparam}}
\newcommand{\closeline}{\lineptslope{\optdist}{\optloss}{\param}}

\newcommand{\unifhat}{\hat \unif}
\newcommand{\paramratio}{\beta}
\newcommand{\critparamratio}{\hat \paramratio}

{\makeatletter
 \gdef\xxxmark{%
   \expandafter\ifx\csname @mpargs\endcsname\relax 
     \expandafter\ifx\csname @captype\endcsname\relax 
       \marginpar{xxx}
     \else
       xxx 
     \fi
   \else
     xxx 
   \fi}
 \gdef\xxx{\@ifnextchar[\xxx@lab\xxx@nolab}
 \long\gdef\xxx@lab[#1]#2{{\bf [\xxxmark #2 ---{\sc #1}]}}
 \long\gdef\xxx@nolab#1{{\bf [\xxxmark #1]}}
}

\begin{document}

\title{Wireless coverage prediction via parametric shortest paths}



\author{David Applegate}
\affiliation{
	\institution{Google Inc., USA}
}
\email{dapplegate@google.com}

\author{Aaron Archer}
\affiliation{
	\institution{Google Inc., USA}
}
\email{aarcher@google.com}

\author{David S. Johnson}
\affiliation{
	\institution{(deceased)}
}

\author{Evdokia Nikolova}
\affiliation{
	\institution{University of Texas at Austin, USA}
}
\email{nikolova@austin.utexas.edu}

\author{Mikkel Thorup}
\affiliation{
	\institution{University of Copenhagen, Denmark}
}
\email{mikkel2thorup@gmail.com}

\author{Ger Yang}
\affiliation{
	\institution{University of Texas at Austin, USA}
}
\email{geryang@utexas.edu}

\renewcommand{\shortauthors}{D. Applegate et al.}


\begin{abstract}
  When deciding where to place access points in a wireless network, it is useful
  to model the signal propagation loss between a proposed antenna
  location and the areas it may cover.  The indoor dominant path (IDP) model,
  introduced by W\"{o}lfle et al., is shown in the literature to have good
  validation and generalization error, is faster to compute than competing
  methods, and is used in commercial software such as WinProp, iBwave Design,
  and CellTrace.  Previously, the algorithms known for computing it involved a
  worst-case exponential-time tree search, with pruning heuristics to speed it
  up.

  We prove that the IDP model can be reduced to a parametric shortest path
  computation on a graph derived from the walls in the floorplan.  It therefore
  admits a quasipolynomial-time (i.e., $n^{O(\log n)}$) algorithm.  We also give
  a practical approximation algorithm based on running a small constant number
  of shortest path computations.  Its provable worst-case additive error (in dB)
  can be made arbitrarily small via appropriate choices of parameters, and is
  well below 1dB for reasonable choices.  We evaluate our approximation
  algorithm empirically against the exact IDP model, and show that it
  consistently beats its theoretical worst-case bounds, solving the model
  exactly (i.e., no error) in the vast majority of cases. 
\end{abstract}



\maketitle

\section{Introduction}
\label{sec:intro}

\begin{figure}
  \includegraphics*[width=0.5\linewidth,keepaspectratio=true]{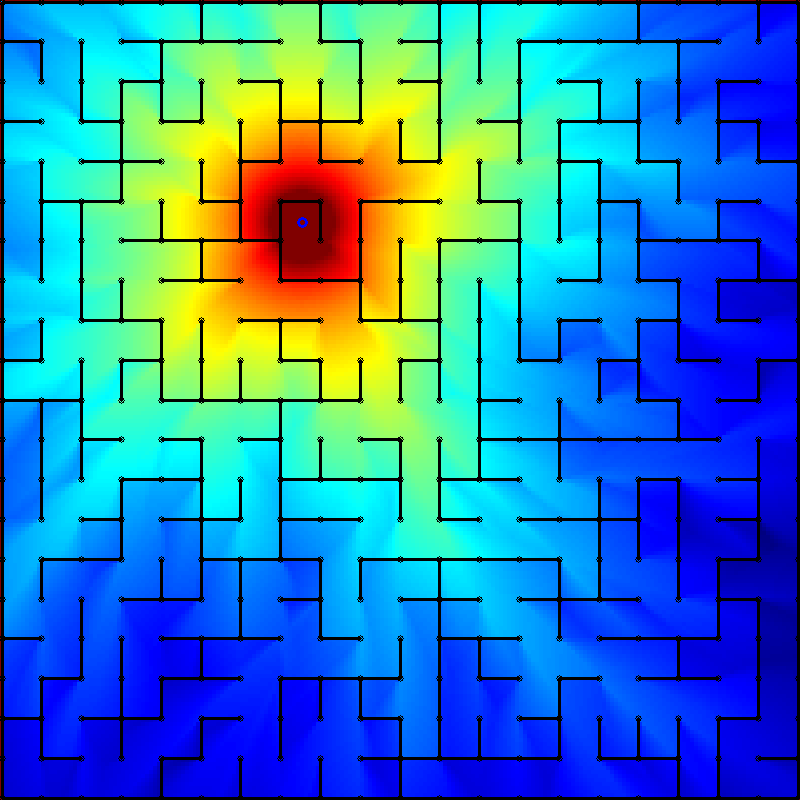}
  \caption{\label{fig:heatmap}Heat map for a random maze.}
\end{figure}

When installing a wireless network in an office or other building, it
can be difficult to figure out the best spots to place access points (APs)
in order to achieve the desired signal strength throughout the
building.  Wireless signal propagation is complicated, especially in
an indoor office environment where the numerous walls attenuate the
signal as it passes through them or diffracts around corners.  Since
physical trial-and-error is an expensive and time-consuming process, it would
be great if we had an effective model to answer the following
question: if we place an access point at location $\src$, how strong of
a signal will we receive at various other points $\dest$ throughout the
building?  In other words, we would like to be able to produce a heat
map such as in Figure~\ref{fig:heatmap}.  If we can answer this
question quickly and accurately in simulation, then it opens the door
to many algorithmic approaches for designing the wireless network to
provide the necessary coverage, throughput, etc. that we desire.

Our original motivation for this paper came several years ago, when one of our
colleagues told us that he had discovered a beautifully simple but heretical
indoor radio signal propagation model in the literature, that was giving him
results that accorded with reality surprisingly well.  The model has two
controversial features: it uses only the strongest propagation path to estimate
the received signal strength at a point, and it completely ignores paths that
rely on reflections off of walls, focusing on diffractions around corners as the
only mechanism for a path to change direction.  This \emph{indoor
  dominant path} (\idp) model lies in stark contrast to the methods blessed by
conventional wisdom in the field: sophisticated ray-tracing techniques that
expend a large amount of computational effort on tracing individual rays as they
bounce off of multiple obstacles, and add up the contributions of multiple rays
at each prediction point, accounting for phase shifting of the waves from the
differential path lengths and the resulting constructive or destructive
interference.  Nevertheless, another member of his team was spending his days
pushing a cart around the halls, measuring the actual received signal strengths
from the WiFi APs in our building, and his team was finding that the
\idp model matched reality roughly as well as much more sophisticated and
accepted ray tracing models.

Better yet, the \idp model is fast: the commercial software he was using that
featured the \idp model could compute a heat map for our entire office building
from a single AP in minutes, many times faster than more sophisticated ray
tracing tools.  However, O(1min) was still not fast enough to satisfy our
colleague.  For each possible AP location on a 1m grid, he wanted to
compute a heat map for the entire building, also at a 1m resolution, in order to
inform his AP placement.  For a 60m x 60m building, this would be 3600 heat maps.
At 1min apiece, this would take 2.5 days of computation.  Obtaining
the relevant data to fuel this modeling was a chore unto itself, and he wanted
to rerun the models as the input data improved.  Could we compute the model
faster, he asked?

Unfortunately, we were not able to do so in time to help our colleague with his
project, but we did subsequently design algorithms to compute the \idp model
faster, and this paper is the result.  On a synthetic 60m x 62m instance meant
to model an office building (Section~\ref{sec:data}), our algorithm takes
roughly 1.3sec per heat map, preceded by 2.9sec of pre-computation
(which can be amortized across all of the heat maps).  Using this algorithm, the
full set of heat maps could be computed in under 80min, despite the fact that we
have taken no particular care to produce an optimized implementation.

The main algorithmic insight of this paper is that we can reduce the \idp model
to a parametric shortest path computation on an associated graph.  By exploiting
further structure, we can reduce this to the equivalent of about 2 ordinary,
non-parametric shortest path computations on the same graph, while incurring an
approximation error that is provably tiny in the worst case, and nearly always zero in
practice.  We call this our \emph{geometric progression} (GP) algorithm, as it
involves evaluating a geometric progression of parameter values, on
geometrically increasing subsets of the original graph.  Our GP algorithm
possesses two benefits over previous algorithms for the \idp model.  First, it
relies on fast polynomial-time algorithms for shortest path (\eg Dijkstra's
algorithm).  Second, it computes path losses for \emph{all measurement points
  simultaneously}.  In contrast, previous algorithms for the \idp model used a
worst-case exponential tree search to compute each point-to-point path loss (with
pruning heuristics to speed it up), and had to do one such computation per
measurement point, rather than handling them all at once.

Plets et al. published a careful study validating the \idp model,
reporting superb agreement between \idp and empirical measurements on four
distinct buildings with diverse physical
characteristics~\cite{DBLP:journals/ejwcn/PletsJVTM12}.  Nevertheless, it is
probably fair to say that the \idp model has not reached widespread acceptance
within the academic community.  Within industry, it has gained more traction,
being prominently featured in at least three commercial software packages for
wireless signal propagation: WinProp~\cite{WinProp-website}, iBwave
Design~\cite{iBwave-Design-press-release}, and
CellTrace~\cite{CellTrace-website}.  WinProp, associated with the original
inventors of the indoor, outdoor, and mixed dominant path models, counts many
large telecom companies and device manufacturers among its customers, including
Alcatel, British Telecom, Ericsson, France Telecom, Fujitsu, Intel, Italtel,
Kyocera, Motorola, Nokia, Nokia Networks, Sony, Swisscom, T-Mobile, and
Vodafone~\cite{WinProp-customers-via-wayback}.\footnote{This customer list
  disappeared from WinProp's website after its acquisition by
  Altair~\cite{altair-acquires-AWE-press-release}, but we reconstructed it
  from the raw HTML in an archived
  snapshot~\cite{WinProp-customers-via-wayback}.}  In particular, Qualcomm has
proved its capability of modeling femtocell performance in urban neighborhoods
\cite{grokop2010simtown, meshkati2009mobility}, and Nokia has used it for
modeling LTE multimedia broadcast systems \cite{awada2016field}. One practical
selling point of the \idp model is its insensitivity to the finer details of a
building's layout~\cite{AschrafiWLLWW06}.  
Most ray tracing tools require CAD
drawings or other detailed databases describing the geometry of a building,
which can sometimes be impossible or prohibitively expensive to obtain.  In
contrast, tools like WinProp can often generate reasonable predictions based on
as little as a photocopy of a floorplan, along with information on the materials
that compose each wall.

Although the \idp model may not yet be fully accepted in the academic community,
we take its commercial success and the strong validation results cited above as
convincing indicators that it merits further study.  While the community would
probably value further validation of the model, that is not the aim of this
paper.  Here, we take the quality of the model as a given, and our goal is to
present a new algorithmic approach that can solve it faster.  The practical
value of such a speedup is to enable new use cases such as the one described
above, where our colleague wished to compute a separate heat map for each
possible transmitter location in a 1m grid, to serve as input for a WiFi network
planning tool.

{\bf Our Contribution.}  This paper provides faster computation of the \idp
model, with rigorous approximation guarantees in worst-case polynomial running
time (Section~\ref{sec:geom-prog}), in contrast to the worst-case exponential
time of the existing approaches.
%
Our \emph{geometric progression} ($GP$) algorithm is also very fast in practice.
While the solution it returns may (rarely) be suboptimal, it is guaranteed to be
very close to optimal.  The algorithm is based on a careful geometric design and
analysis related to the shape of the nonlinear objective function when projected
on a 2D subspace spanned by the distance and loss parameters in our problem. For
reasonable parameter settings, the worst-case additive error is a fraction of
1dB (Theorems~\ref{thm:worst-error} and \ref{thm:expected-error}), and our
computational experiments (Section~\ref{sec:experiments}) show that it actually
solves the \idp model exactly (\ie with no error) in the vast majority of cases.
Under reasonable assumptions, the total running time of the $GP$ algorithm is
dominated by a single (not parametric) shortest path computation on a graph
derived from the corner points of the walls in the floorplan.
To sum up, $GP$ is a fast, practical algorithm with tiny worst-case
error bounds that are essentially zero in practice.  

In addition, we provide a (slower but still fast) exact algorithm
(Section~\ref{sec:st-hull}), which we use to evaluate the approximation errors
of the GP algorithm (Section~\ref{sec:experiments}).
%
Our exact algorithm enumerates feasible paths that
correspond to optimal paths in the parametric shortest path problem.
The parametric problem takes two weights per edge, a
distance $\dist_e$ and loss $\loss_e$, and asks what are the shortest
paths with respect to edge weights $\loss_e +
\param \dist_e$, for all values of the parameter $\param \in [0,
\infty]$.  The parametric complexity refers to the number of distinct such
shortest paths.  The values of $\param$ for
which the shortest path switches from one path to another are
called \emph{breakpoints}.  Our exact algorithm runs in time
proportional to the number of breakpoints, so bounds on this quantity
are of interest.


Carstensen constructed examples where the number of breakpoints is
$n^{\Omega(\log n)}$~\cite{carstensen}.  Although our exact algorithm
for \idp suffers from this worst-case lower bound, we do show that it
has polynomial-time smoothed complexity (Theorem~\ref{thm:smooth}), meaning that the bad examples
with $n^{\Omega(\log n)}$ breakpoints are pathological and fragile.
In our experiments, the average number of breakpoints is roughly 4 to
6 (Section~\ref{sec:approx-errors}).  This means that our exact
algorithm requires about 5 shortest path computations per
measurement point that we wish to model, compared to GP running the
equivalent of $O(1)$ shortest path computations to closely approximate the
model for \emph{all} measurement points \emph{simultaneously}.




{\bf Additional Related Work.} 
W\"{o}lfle and Landstorfer first introduced the dominant path model in
1997~\cite{WolfleLandstorfer-97-orig-dompath}.  They suggested solving the model
approximately (with no stated guarantees) using a heuristic grid-based method
reminiscent of fast marching methods~\cite{Sethian99-fast-marching-book}.
Shortly, the same authors suggested a different algorithm based on
searching an exponential-sized tree, with pruning heuristics to speed it
up~\cite{WolfleLandstorfer-98-tree-based}.  In 1998, they also founded a company
called AWE Communications to develop a highly-successful software tool called
WinProp, enabling wireless coverage prediction and network design.  AWE
was acquired by Altair in April 2016 \cite{altair-acquires-AWE-press-release}.
AWE and affiliated researchers issued a blizzard of similar conference and
workshop papers, so it is difficult to choose which ones to cite,
but~\cite{Wolfle05-IDP} seems to be a canonical choice for the IDP model.

Multiple papers (\eg
\cite{WolfleLandstorfer-98-tree-based,Wolfle05-IDP,DBLP:journals/ejwcn/PletsJVTM12,awada2016field,grokop2010simtown,meshkati2009mobility})
compare predictions of the dominant path model against actual physical
measurements.  The one by Plets \etal~\cite{DBLP:journals/ejwcn/PletsJVTM12}
stands out for its careful methodology and exposition.  They report superb
agreement between the \idp model and their empirical measurements on four
distinct buildings with diverse physical characteristics, with a mean
absolute modeling error of 1.29 dB on the best building
and 3.08dB on the worst.

%
In the algorithms literature, Gusfield gave an upper bound of $n^{O(\log n)}$ on the complexity of the parametric shortest path
problem~\cite{Gusfield:1980:SAC:909661}. 
Nikolova \etal~\cite{nikolova-ssp} gave an exact algorithm based on parametric
shortest paths for a stochastic routing problem of maximizing the probability of
arriving on time.  Nikolova~\cite{nikolova-approx} later gave approximation
algorithms for a general combinatorial optimization framework with several
concave objective functions.  At a high level, the approximation algorithms use
geometric progressions similarly to our main algorithm here, but they do not
apply to our problem due to a difference in objective functions and desired bounds.  To
wit, those algorithms provide multiplicative approximations while our algorithm
gives an additive approximation requiring a different geometric analysis.  In
our context, a multiplicative approximation is meaningless, since path losses
are measured on a log scale (dB), so the units would not even make sense for a
multiplicative error.  Our algorithm also provides a better tradeoff between
the error and number of shortest path invocations.
It remains an open question whether
there exists a polynomial-time algorithm to solve the \idp exactly,
but Carstensen's lower bound implies that we cannot hope to do so
directly via our reduction to parametric shortest paths.



\section{Dominant path model and reduction}
\label{sec:model}

Ideal freespace isotropic radio signal propagation is simple: the energy
received by an antenna is proportional to $1/\dist^2$, where $\dist$ is the
distance from the transmitter to the receiving antenna.  Real-world radio
signals are non-isotropic, diffract around corners, and can penetrate through
or reflect off of walls.  The dominant path model is motivated by a simple
empirical observation: although multiple propagation paths can all contribute to
the received signal strength at a measurement point $\dest$, usually the top 2
or 3 rays account for more than 95\% of the energy\cite{Wolfle05-IDP}.
Therefore, predicting the single path with smallest propagation loss can drive a
good model of the total received signal strength.  Even supposing that the
dominant path contributes only 50\% of the total energy, this would induce only
a 3dB error, which is broadly within the range of accuracy that one achieves
with more sophisticated models.  The study by Plets et al. validates the IDP
model that uses just the single dominant path, reporting average errors within
this range.

\begin{figure}
  \includegraphics*[width=0.5\linewidth,keepaspectratio=true]{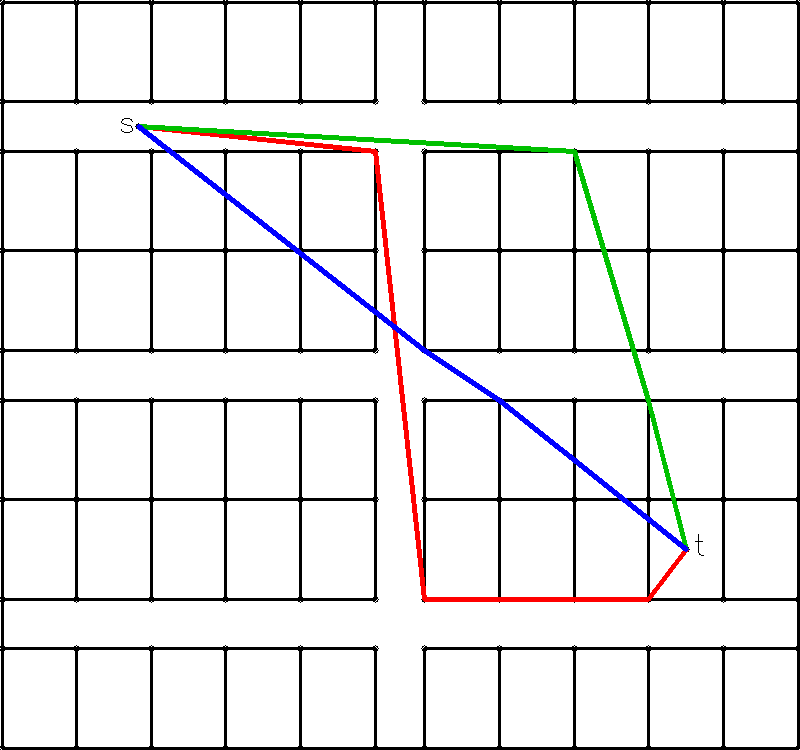}
  \caption{\label{fig:paths}Paths on convex hull for office building}
\end{figure}

The indoor dominant path model focuses on RF propagation inside
buildings, such as for WiFi.  It models the path loss $\PL(\Path)$
along any path $\Path$, finds the path $\Path$ from source (\ie
transmitter location) $\src$ to destination $\dest$ that minimizes
$\PL(\Path)$, and uses that number for the propagation loss from
$\src$ to $\dest$.  In this minimization, it considers only polygonal
paths that change direction only at corner points of walls in the
floorplan.  Figure~\ref{fig:paths} shows examples of such paths, on a
synthetic floorplan representing a generic office building.

The path loss $\PL(\Path)$ along path $\Path$ can be broken into four
components: (1) the (constant) unobstructed path loss at a reference
distance $\dist_0$ (typically 1m), (2) a distance term based on the
ratio $\dist / \dist_0$, (3) penetration losses for passing through
walls, and (4) diffraction losses for changing directions around
obstacle corners.  With the terms in this order, the
formula for the path loss $\PL(\Path)$ (measured in dB) is:
\begin{equation}
  \label{eq:path-loss-model}
  \PL(\Path) = \PL_0 + 10 \fsexp \log_{10} \frac{\dist(\Path)}{\dist_0} +
  \sum_i \wallloss_i + \sum_j \diffracloss_j
\end{equation}
where the path $\Path$ intersects some sequence of walls $i$ and
changes directions at some sequence of corners $j$.  Since $\PL_0$ is
a constant (40 dB at 1m~\cite{DBLP:journals/ejwcn/PletsJVTM12}), we
ignore it for the rest of this paper, until the experiments.

We assume throughout that the frequency of the signal is fixed.  The
example parameter values we state here pertain to 2.4 GHz.  The wall
penetration loss term $\wallloss_i$ is modeled as a constant for each
wall, accounting for material composition and thickness (\ie
$\wallloss_i$ is given as input).  Plets
\etal~\cite{DBLP:journals/ejwcn/PletsJVTM12} cite some typical values
for thin walls, such as 2dB for glass or layered drywall, 6dB for
wood, 7dB for brick, and 10dB for concrete, and for thick walls, 4dB
for glass and 15dB for concrete.  Note that the modeled wall
penetration loss does not depend on the angle at which the path
intersects the wall.
The diffraction loss at corner $j$ is modeled as
the deflection angle $\Angle_j$ times some constant $\diffconst_j$
that depends on the wall material at corner $j$.  Plets et al. use
$\diffconst_j = 5\text{dB} / 90^{\degree}$ for layered drywall, and
$17.5\text{dB} / 90^{\degree}$ for concrete walls, which they
determined empirically \cite{DBLP:journals/ejwcn/PletsJVTM12}.

Although earlier works (\eg~\cite{Wolfle05-IDP}) used various values
of $\fsexp > 2$ that were tuned to fit measurements on specific
buildings, Plets \etal recommend $\fsexp = 2$, corresponding to
isotropic freespace loss, because it agrees well with their
experiments and doesn't require tuning, while being simple and easy to
justify theoretically.  Therefore, we also use $\fsexp = 2$ in our
computational experiments, although our theorems work for every
choice of $\fsexp$.

Critically, the dominant path model does \emph{not} account for
reflections.  Some versions of the model incorporate other effects,
such as so-called ``waveguiding'' along tunnels or
corridors~\cite{Wolfle05-IDP}.  Following Plets \etal again, we eschew
these extra knobs.

The heatmap in Figure~\ref{fig:heatmap} shows the result of solving
this model for a synthetic building meant to represent a maze (see 
Section~\ref{sec:data} for details).  Discontinuities at each wall are
obvious.  A closer look also reveals ``shadows'' behind each corner,
as the diffraction loss gradually accumulates for points whose dominant
path bends around the corner.

Previous algorithms for solving the dominant path model relied on a
tree search that could take exponential time in the worst case.  The
central theoretical contributions of this paper are to reduce the
dominant path problem to a parametric shortest path problem, then
give practical algorithms with rigorous guarantees to solve it.

\subsection{Parametric shortest paths}
\label{sec:parametric-SP}
This section defines the parametric shortest path problem in graphs,
and the next section reduces the dominant path problem to it.

Our input is a graph $\graph = (\nodes, \edges)$ where $\nodes$ is a
set of nodes and $\edges$ is a set of edges, along with two
non-negative weights on each edge $e$: a distance $\dist_e$ and a loss
$\loss_e$.  In our context, $\dist_e$  represents Euclidean
distance and $\loss_e$ represents penetration and diffraction
loss.  Given any parameter $\param \geq 0$, we  define a \emph{hybrid}
edge weight $\Hparam(e) = \loss_e + \param \dist_e$.

Fix a source node $\src \in \nodes$ and target node $\dest \in
\nodes$, and imagine computing the shortest \st path $\Pparam$ in
$\graph$ with respect to edge weights $\Hparam$, as $\param$ sweeps
from 0 to $\infty$.  Note that $\Path_0$ is simply the shortest path
according to loss vector $\loss$, whereas $\Path_\infty$ 
is the shortest path according to distance vector $\dist$.

Let $\dlparam$ denote the loss and distance of path $\Pparam$.
There are only a finite number of possible paths, so as $\param$
increases from 0 to $\infty$, $\Pparam$ remains constant for some
$\param$ interval, then switches to another path at some \emph{critical
  value} of $\param$, also known as a \emph{breakpoint}.  The
\emph{parametric \st shortest path problem} is to compute the full set
of distinct paths $\{\Pparam : \param \in \zeroinfty\}$, or, if
we care only about their weights, then the full set of distance-loss
pairs: $\{\dlparam : \param \in \zeroinfty \}$.  The \emph{parametric single-source
  shortest path problem} is to solve the parametric \st shortest path
problem 
for a single $\src$ and all $\dest$.

Back to the \st version, as $\param$ increases through a critical
value, we always trade off a higher loss for a smaller distance.  That
is, $\lparam$ increases with $\param$, and $\dparam$ decreases with
$\lambda$.  We will refer often to $\dist_0$ and $\dist_\infty$.  To
avoid confusion over the fact that $\dist_0 \geq \dist_\infty$, we
introduce the aliases $\dmax := \dist_0$ and $\dmin := \dist_\infty$.
Usually, a fixed destination $\dest$ will be clear from context so we
suppress $\dest$ in the notation as we have done above, but
technically $\Pparam(t)$, $\dparam(t)$, $\lparam(t)$, $\dmin(\dest)$
and $\dmax(\dest)$ do depend on $\dest$.

\begin{figure}
  \begin{center}
    \includegraphics*[width=3.3in,keepaspectratio=true,clip=true,trim=1.53in 1.22in 1.6in 2.38in]{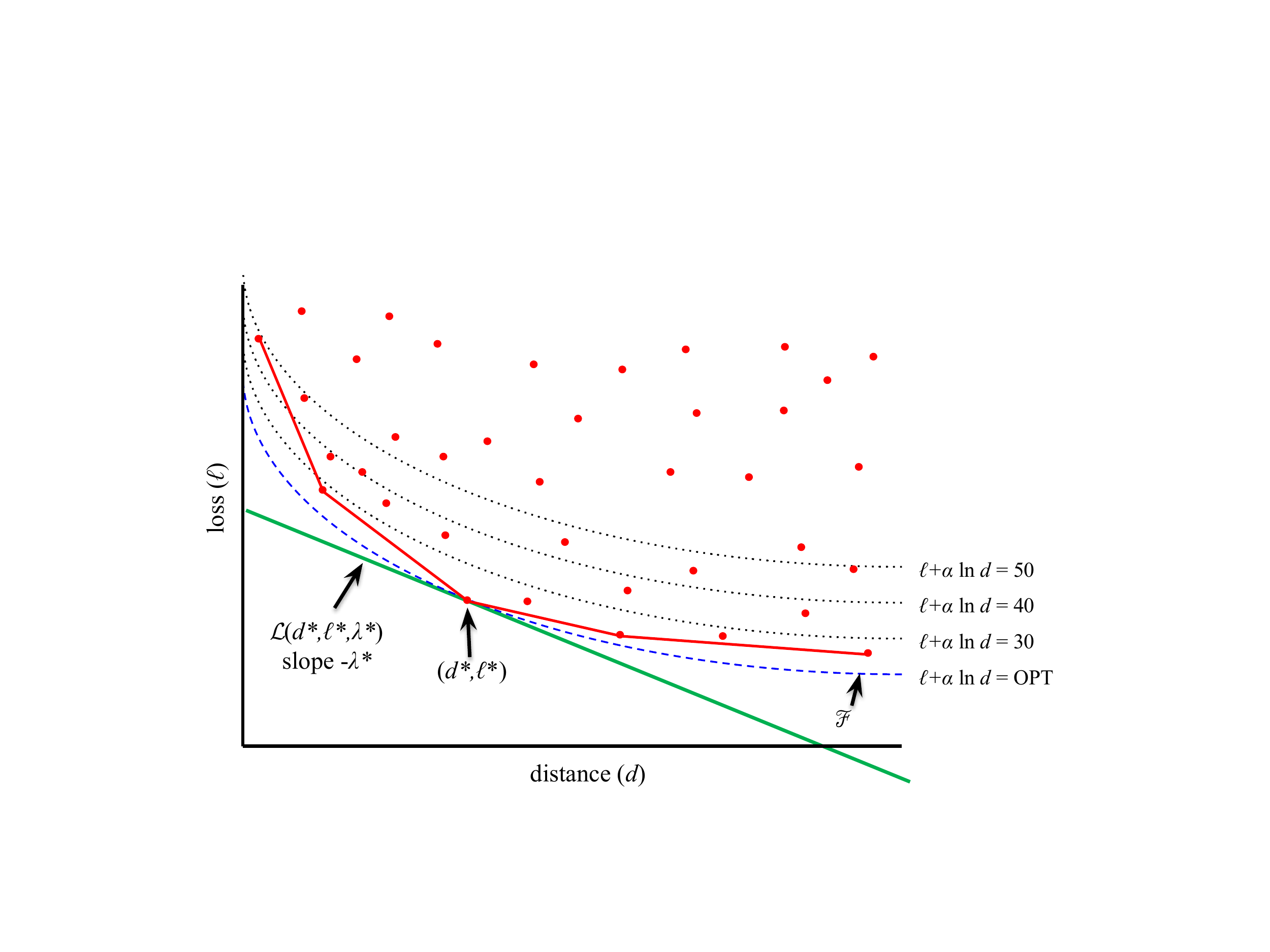}
  \end{center}
  \caption{\label{fig:level-sets-cvx-hull}Level sets and convex hull}
\end{figure}

Figure~\ref{fig:level-sets-cvx-hull} visualizes what it means
geometrically to minimize the function $\Hybrid{\param}$ over all \st
paths.  For each of the exponentially-many \st paths $\Path$, imagine
plotting the point $(\dist(P), \loss(P))$ in $\R^2$.  Let
$\pointcloud$ denote this cloud of points (\ie the red points in 
Figure~\ref{fig:level-sets-cvx-hull}).  Each level set of the function
$\Hybrid{\param}$ is a line of slope $-\param$.  Place a line of slope
$-\param$ below the point cloud, and move it up until it first bumps
into one of the points.  This point corresponds to the path $\Path$
that minimizes $\Hybrid{\param}(\Path) = \loss(\Path) + \param
\dist(\Path)$.  If we start with $\param$ sufficiently high, we hit
$(\dmin, \loss_\infty)$.  As we lower $\param$, the optimal level set
line rotates around $(\dmin, \loss_\infty)$ until it hits another
point in the cloud.  At this value of $\param$, two paths are tied as
the shortest \wrt edge weights $\Hybrid{\param}$.  As $\param$
decreases further, we rotate the line about this new point, until we
hit a third point, and so on.  By the time $\param$ drops to 0, we
have constructed the lower-left convex hull of our point cloud, ending
at $(\dmax, \loss_0)$ (\ie the red polyline in
Figure~\ref{fig:level-sets-cvx-hull}).

The corner points (\extremepts) of this convex hull correspond to the paths output
by the parametric shortest path computation.  To avoid confusion with the corners of walls, we will henceforth refer to the corners of the convex hull as \emph{\extremepts}.  The three paths shown in
Figure~\ref{fig:paths} are in fact paths that correspond to \extremepts for the \st pair shown.  These \extremepts partition the
$\param$ range $\zeroinfty$ into segments, where each \extremept\
represents the shortest path \wrt $\Hybrid{\param}$ for all $\param$
in its segment.  The boundary points between these segments represent
the critical values of $\param$, which correspond to (the negative of)
the slopes of the red line segments in
Figure~\ref{fig:level-sets-cvx-hull}.  These facts will be important
for understanding both our geometric progression algorithm
(Section~\ref{sec:geom-prog}) and some of our experimental results
(Section~\ref{sec:experiments}).

\subsection{A graph representing all valid paths}
\label{sec:reduction}
This section constructs a graph with two weight functions $\dist$ and
$\loss$ that capture the distance and (penetration loss + diffraction
loss) of paths in the dominant path model.  We must define a graph
$\graph$ and edge weights $\dist_e$ and $\loss_e$ such that:
\begin{enumerate}
\item the distance $\dist(\Path)$ and loss $\loss(\Path)$ of path
  $\Path$ in the dominant path model equal $\sum_{e \in \Path}
  \dist_e$ and $\sum_{e \in \Path} \loss_e$, and
\item every valid physical path $\Path$ in the dominant path model
  corresponds to a path in the graph,
\item every undominated path in the graph corresponds to a physical
  path that is considered in the dominant path model.
\end{enumerate}
We construct such a graph in two phases: first a graph $\gOne$ that
encodes all relevant paths, plus their distances and wall penetration
losses, then a related graph $\gTwo$ that also encodes diffraction
losses.

Constructing $\gOne = (\nodes_1, \edges_1)$ is straightforward.  The
node set $\nodesOne$ is the union of three sets: the single
source $\{\src\}$, the set of $\dests$ \emph{destinations}  (\aka
\emph{measurement} points), and the set of \emph{corners}
$\corners$.  Here, $\dests$ means the set of all points for which we
wish to compute the dominant path, and $\corners$ means the set of
endpoints of wall segments in the floor plan.  The set
$\edgesOne$ is the union of four sets of directed edges: all
$\src-\corners$, $\src-\dests$, $\corners-\corners$, and
$\corners-\dests$ pairs.
There is no need for $\dests-\dests$ edges, since all intermediate
points in the polygonal \st paths considered in the dominant
path model are corner points.
For each $e \in \edges_1$, set $\dist_e$ to the Euclidean distance between
its endpoints, and $\loss_e$ to the sum of the penetration losses of
all walls it crosses.

\begin{figure}
  \begin{center}
    \includegraphics*[width=0.8\linewidth,keepaspectratio=true,clip=true,trim=1.10in 2.46in 3.67in 1.24in]{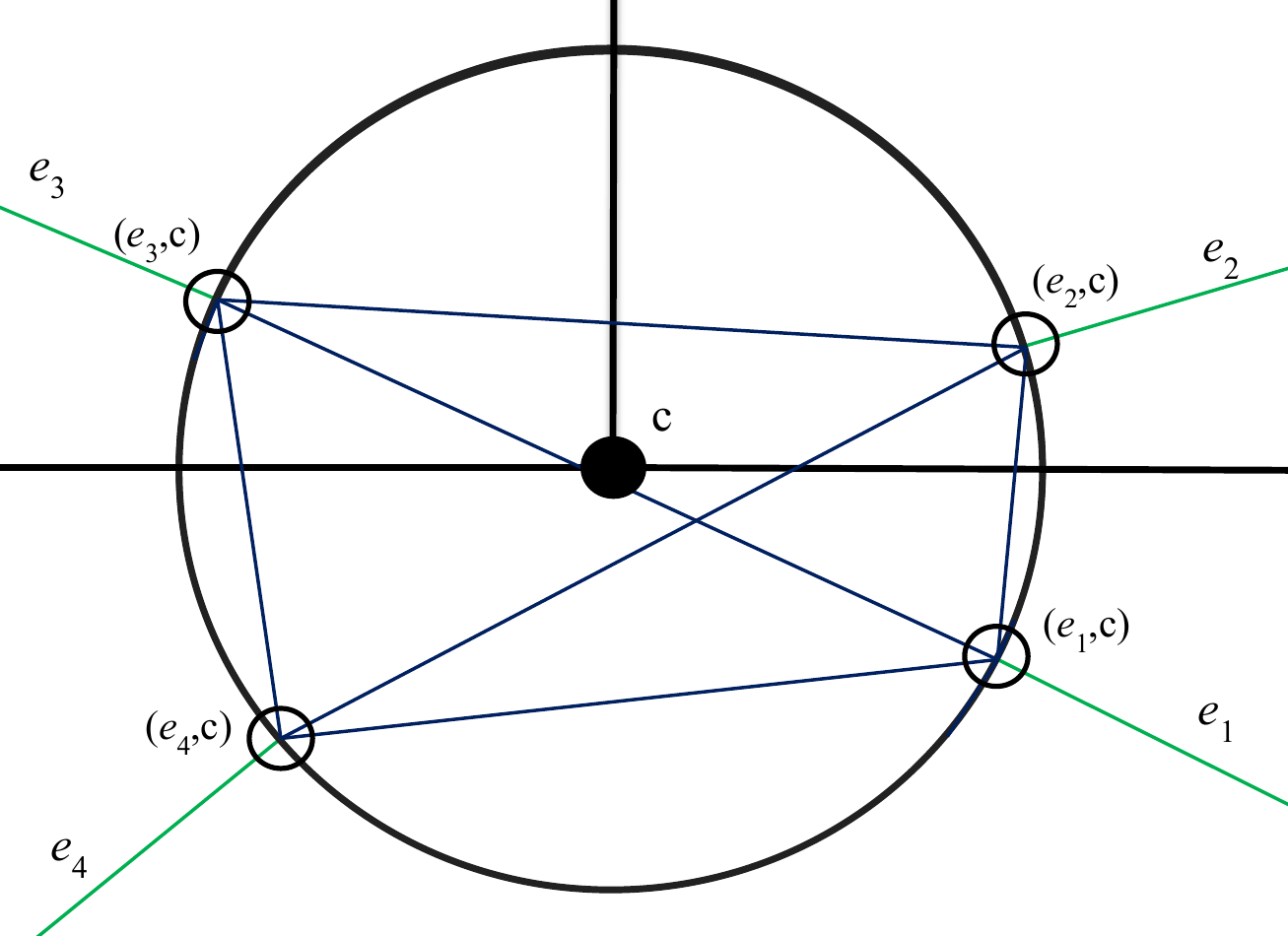}
    \end{center}
  \caption{\label{fig:G2-corner}G2 corner}
\end{figure}

There are two defects in $\gOne$ that we must correct in our
construction of $\gTwo$: it models neither the penetration losses at
corners nor the diffraction losses.  We solve this problem by
``exploding'' each corner node $\corner \in \corners$, replacing it
with a set of new nodes, one for each incoming and outgoing edge $e$.
These new nodes are \emph{directed sockets} of
$\gOne$, \ie ordered pairs $(e,\corner)$ where $e \in \edgesOne$ is
incident to $\corner$ in $\gOne$.  These new socket nodes are
illustrated by the circles in Figure~\ref{fig:G2-corner}, labeled
$(e_1,\corner),\ldots,(e_4,\corner)$. We now add a directed edge from
each incoming socket to each outgoing socket at corner $\corner$,
represented by the six blue edges inside the big circle in
Figure~\ref{fig:G2-corner}.

For a new intra-cluster edge $e \in \edgesTwo$ 
that runs from incoming socket $(e_1, \corner)$ to outgoing socket $(e_2, \corner)$, 
$\dist_e = 0$ as 
this edge covers zero physical distance.
The loss $\loss_e$ is the sum of two terms: one for the diffraction
loss, and one for the penetration loss.  The diffraction loss is
$\diffconst_{\corner} \Angle_{e_1 e_2}$, where $\Angle_{e_1 e_2}$
is the physical angle between directed edges $e_1$ and $e_2$.  For the
penetration loss, we consider the total loss for the walls incident at
corner $\corner$ encountered as we sweep either clockwise or
counterclockwise from $e_1$ to $e_2$, and take the minimum.

There are two more subtleties.  First, there will be edges of
$\gOne$ that run from one end of a wall segment to the other.  We must
actually represent these as two edges, one on each side of the wall,
so that we can compute the penetration losses correctly in the
intra-corner edges of $\gTwo$.  Second, sets of co-linear corner\
points are a common occurrence in buildings, so we cannot assume away
their existence.  It would be problematic to consider edges directly from one
end of the line of corners to the other, because we would have to make
a decision at each intermediate corner point about which side of the
wall the edge lies on for that segment, thereby introducing an
exponential number of parallel edges.  Instead, we simply
delete these edges, keeping only the edges
connecting adjacent pairs along the line of corner points.

It is clear that every valid path $\Path$ in the dominant path model
is represented by a corresponding path in $\gTwo$, and $\dist(\Path)$
and $\loss(\Path)$ as defined by $\gTwo$ agree with the values
assigned by the dominant path model.  Conversely, every path in
$\gTwo$ that visits each corner point at most once corresponds to a
valid path.  There also exist paths $\Path$ in $\gTwo$ that
visit a corner point more than once via different sockets from
$\gOne$, which are technically different nodes in $\gTwo$ and hence
do not violate the definition of a path being a trail with no cycles.
However, it is easily seen that those paths are always dominated by
paths that do not revisit corners.

To sum up, for each path $\Path$ considered valid by the dominant
path model, there is a corresponding path $\Path$ in $\gTwo$, and we
have set the edge weights in $\gTwo$ such
that~\eqref{eq:path-loss-model} becomes
\begin{equation}
  \label{eq:PL-in-G2}
  \PL(\Path) = \PL_0 + \loss(\Path) + 10 \fsexp \log_{10}
  \dist(\Path)
\end{equation}
Moreover, all non-valid paths in $\gTwo$ are irrelevant.  To make this
mathematically cleaner, we transform the $\log_{10}$ to $\ln$, plug in
the freespace loss value of $\fsexp = 2$, and define $\logmult =
\frac{10 \fsexp}{\ln 10} \mathrm{dB} \approx 8.686 \dB$.  We also
define $\obj(\dist,\loss) = \loss + \logmult \ln \dist$, and
$\obj(\Path) = \obj(\dist(\Path), \loss(\Path))$.  With these
transformations, the dominant path is the path $\Path$ that minimizes $\obj(\Path)$.

\subsection{Reduction to parametric shortest path}
This section demonstrates that the dominant \st path is one of the
parametric shortest \st paths in $\gTwo$.

Suppose $\optP$ is the dominant \st path, and  define
$\optdist = \dist(\optP)$, $\optloss = \loss(\optP)$, and $\optparam =
\logmult / \optdist$.  We devote the rest of this section to proving
the following theorem:

\begin{thm}
  \label{thm:reduction}
  Define $\optdist = \dist(\optP), \optparam =
  \frac{\logmult}{\optdist}$ where $\optP$ is the dominant \st path.
  Then $\optP$ is also the $\Path$ that minimizes
  $\Hybrid{\optparam}(\Path) = \loss(\Path) + \optparam \dist(\Path)$.
\end{thm}

We introduce some notation to set up the proof.  The solution to the
parametric \st shortest path problem yields $\{ \Pparam : \param \in
\zeroinfty \}$ and their corresponding distance-loss pairs $\{
\dlparam : \param \in \zeroinfty \}$.  These are actually finite sets,
one per critical value of $\param$, so we can simply evaluate $\loss +
\logmult \ln \dist$ for each of these $(\dist, \loss)$ pairs, and
select the smallest one.  Since $(\dist_\optparam, \loss_\optparam)$
is one of these pairs, and Theorem~\ref{thm:reduction} says that it is
the dominant path, this solves the dominant path model.

To finish the reduction, we just need to prove
Theorem~\ref{thm:reduction}.  This is surprisingly simple, as
illustrated in Figure~\ref{fig:apx-bound}.  First, we define
$\lineptslope{\dist}{\loss}{\param}$ to be the line through $\pair$ of
slope $-\param$.

\begin{myproof}
  As in Section~\ref{sec:parametric-SP}, imagine the point cloud
  $\pointcloud$ of $(\dist(\Path), \loss(\Path))$ pairs, for all paths
  $\Path$.  Let $\optloss = \loss(\optP)$ and $\opt = \obj(\optdist,
  \optloss) = \optloss + \logmult \ln \optdist$ and consider the level
  curve $\levelcurve := \{ (\dist, \loss) : \loss = \opt - \logmult
  \ln \dist \}$ of $\obj$ running through $(\optdist, \optloss)$.
  Because $\optP$ is the optimal path, the rest of $\pointcloud$ lies
  on or above this level curve.  Because $\ln \dist$ is a concave
  function of $\dist$, this level curve is convex, and it therefore
  has a supporting tangent line $\opttangent$ at $(\optdist,
  \optloss)$ of slope $-\optparam = -\frac{\logmult}{\optdist}$ (\aka
  the derivative of the level curve at $\optdist$), which is a level
  set for $\Hybrid{\param}$.  The point $(\optdist, \optloss)$
  minimizes $\Hybrid{\optparam}$ because the $\opttangent$ passes
  through $(\optdist, \optloss)$, the rest of $\levelcurve$ lies
  strictly above the tangent line, and the rest of $\pointcloud$ lies
  strictly above $\levelcurve$.
\end{myproof}

We note that a similar theorem and proof would work if $\ln \dist$ were
replaced by any other concave function of $\dist$.



\section{Exact convex hull}
\label{sec:st-hull}

By the results of Section~\ref{sec:model}, solving the \st dominant
path model reduces to a parametric \st shortest path computation, or
equivalently finding the lower left convex hull of the point cloud $\pointcloud$.
Let $\SP(\param)$ denote the shortest path calculation \wrt edge
weights $\Hybrid{\param}$.  We now prove the following theorem:

\begin{thm}
  \label{thm:st-hull}
  The parametric \st shortest path problem can be solved using $2
  \numBreaks - 1$ \st shortest path computations, where $\numBreaks$
  is the number of \extremepts, including both endpoints.
\end{thm}
\begin{myproof}
  Recall that $\SP(\infty)$ yields the leftmost point in the convex
  hull, $(\dmin, \loss_{\infty})$, whereas $\SP(0)$ yields the
  rightmost point, $(\dmax, \loss_0)$.  Next, set $\param =
  (\loss_{\infty} - \loss_0) / (\dmax - \dmin)$ and compute $\SP(\param)$. One of two things
  happens.  If $(\dmin, \loss_{\infty})$ and $(\dmax, \loss_0)$ are
  the only points on the convex hull, then $\SP(\param)$ will return one of these two
  points, proving that all other points lie on or above the line
  $\lineptslope{\dmin}{\loss_\infty}{\param}$, so we are done.
  Otherwise, it will return some point $(\dist, \loss)$ below this
  line.  In that case, we continue constructing the convex hull
  recursively on each of the two intervals $[\dmin, \dist]$ and
  $[\dist, \dmax]$.  Every value of $\param$ we try either discovers a
  new \extremept or proves that the segment connecting the \extremepts at
  either end of the current interval is indeed
  a facet of the convex hull.  Therefore, the number of shortest path
  computations is $\numBreaks + (\numBreaks - 1) = 2 \numBreaks - 1$.
\end{myproof}

As noted in \cite{Gusfield:1980:SAC:909661}, $\numBreaks$ is at most $n^{O(\log n)}$. Therefore, the parametric \st shortest path problem can be solved using at most $n^{O(\log n)}$ shortest path computations.



\section{Geometric progression}
\label{sec:geom-prog}

From Theorem~\ref{thm:st-hull}, we can compute an \st dominant path
efficiently as long as the number of breakpoints is small.  This
observation drives our smoothed analysis in Section~\ref{smoothed}.
However, if we wish to compute dominant paths from a single source
$\src$ to all destinations then we can do much better, especially if
we are willing to tolerate a small additive approximation error.  Our
\emph{geometric progression} (GP) algorithm does precisely this.  We
devote this section to its definition, then to analyzing its
approximation error and practical time complexity.

Given fixed numbers $\gpratio > 1$ and $\param_0 > 0$, define a
geometric progression of parameter values $\param_i = \param_0
\gpratio^i$ for $i \in \ints$.  It is safe to think of $\gpratio$ as
2.  The geometric progression algorithm $\GPd$ runs $\SP(\param_i)$
for $\param_i$ in some sufficiently wide range (that we specify
later).  For each destination point $\dest$, it then outputs the best
of the $\src$-$\dest$ paths $\Path$ it found, according to the real
objective function $\dpobj$.  Since our algorithm always outputs the
path loss of some valid path, it never underestimates the optimal path
loss.

For each $\dest$, Theorem~\ref{thm:reduction} guarantees there is
some $\param$ such that $\SP(\param)$ finds the dominant \st path.
Although we don't know which $\param$ that is, the $\GPd$ algorithm is
guaranteed to use some nearby value, and this allows us to bound our
error, as shown by Theorems~\ref{thm:error},~\ref{thm:worst-error} and
\ref{thm:expected-error}.

\begin{thm}
  \label{thm:error}
  If $\optpair$ is the optimal distance-loss for destination point
  $\dest$ and $\optparam = \logmult / \optdist$, then $\SP(\param)$
  with $\param = \paramratio \optparam$ yields a path
  loss $\obj(\dparam, \lparam) \leq \obj \optpair + \logmult (-1 + \paramratio -
  \ln \paramratio)$.
\end{thm}

It turns out that $\critparamratio := \frac{\gpratio \ln
  \gpratio}{\gpratio - 1}$ is the worst value of $\paramratio$.

\begin{thm}
  \label{thm:worst-error}
  If $\optpair$ is the distance-loss for the dominant path to
  destination $\dest$, then algorithm $\GPd$ returns an \st path
  $\Path$ with $\obj(\Path) \leq \obj \optpair + \logmult (-1 +
  \frac{\ln \gpratio}{\gpratio - 1} + \ln (\gpratio - 1) - \ln \ln
  \gpratio)$ dB.
\end{thm}

\begin{thm}
  \label{thm:expected-error}
  Set $\param_0 = \gpratio^\unif$ where $\unif$ is chosen randomly
  from the distribution $U(0,1)$.  Then for each destination $\dest$,
  algorithm $\GPd$ returns an \st path $\Path$ with $E[\obj(\Path)]
  \leq \obj \optpair + \logmult (- \half \ln \gpratio +
  \ln(\gpratio - 1) - \ln \ln \gpratio)$ dB, where $\optpair$ is the
  distance-loss of the dominant path.
\end{thm}

Note that the expectation in Theorem~\ref{thm:expected-error} is \wrt
the random choice of $\unif$, whereas the input is assumed to be
adversarial.  As we prove these theorems, some of our intermediate
results will tell us what range of $\param_i$ we must consider, and
also allow us to prune $\gTwo$ before running each $\SP(\param_i)$
calculation.  In particular, each measurement point  appears in
only a small number of these graphs, $O(1)$ in practice.

\begin{figure}
  \begin{center}
    \includegraphics*[width=0.6\linewidth,keepaspectratio=true,clip=true,trim=2.61in 2.05in 3.89in 3.46in]{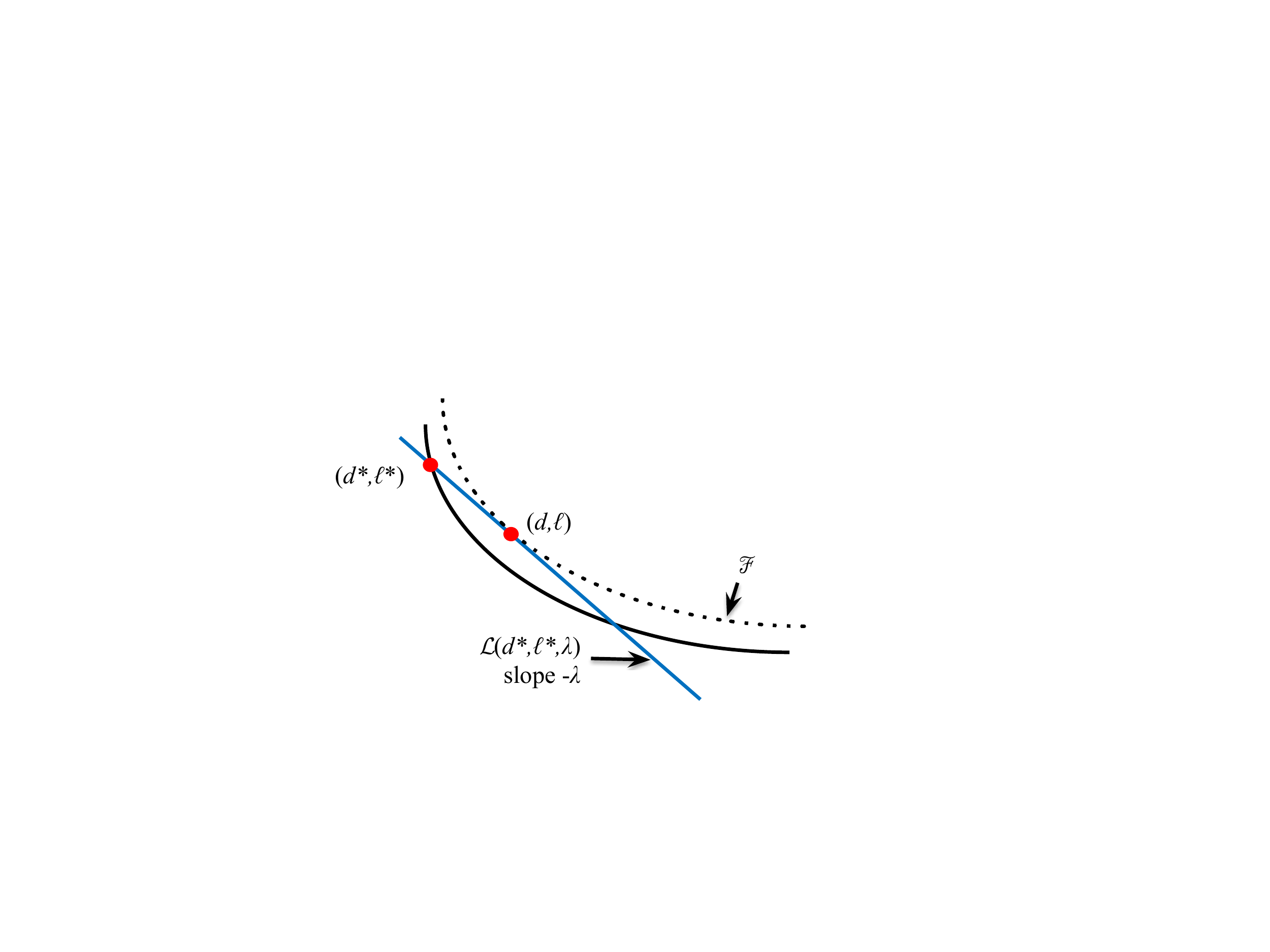}
  \end{center}
  \caption{\label{fig:apx-bound}Approximation bound}
\end{figure}

\begin{proofof}{Theorem~\ref{thm:error}}
  The following geometric argument is illustrated in Figure~\ref{fig:apx-bound}.
  Since $\SP(\param)$ minimizes $\Hparam$, we know $\Hparam \pairp
  \leq \Hparam \optpair$.  Hence, $\pairp$ must lie on or below the
  line $\closeline$, the level set for $\Hparam$ through $\optpair$.
  Among all such points, the one that maximizes $\obj$ occurs at the
  point of tangency between $\closeline$ and some level curve
  $\levelcurve$ of $\obj$.  There is only one such point of tangency,
  which occurs at $\dist = \logmult / \param$, $\loss = \optloss
  - \param (\dist - \optdist)$.  We upper bound $\obj \dlparam$ by
  $\obj \pair$.
  \begin{align*}
    \obj \pairp &\leq \obj \pair = \obj \optpair + (\obj \pair - \obj \optpair) \\
    &= \obj \optpair + (\loss - \optloss) + \logmult \ln (\dist / \optdist) \\
    &= \obj \optpair - \param (\dist - \optdist) - \logmult \ln \paramratio \\
    &= \obj \optpair - \param (\logmult / \param - \logmult / \optparam) - \logmult \ln \paramratio \\
    &= \obj \optpair + \logmult ( -1 + \paramratio - \ln \paramratio)
  \end{align*}
\end{proofof}

\begin{proofof}{Theorem~\ref{thm:worst-error}}
  Let $\optparam = \logmult / \optdist$.  Because consecutive
  $\param_i$ are spaced by a multiplicative $\gpratio$, one of them
  (call it $\paramlo$) must land in the range $[\frac{1}{\gpratio}
  \optparam, \optparam]$, and one (call it $\paramhi$) must land in
  the range $[\optparam, \gpratio \optparam]$.  Thus,
  $\paramhi = \paramratio \optparam$ and $\paramlo =
  \frac{\paramratio}{\gpratio} \optparam$ for some $\paramratio \in
  [1, \gpratio]$.  Applying our upper bound from
  Theorem~\ref{thm:error} twice, we see that the path $\Path$ returned by
  $\GPd$ has error at most
  \[
  \logmult (-1 + \min (\paramratio - \ln
  \paramratio, \paramratio / \gpratio - \ln
  \paramratio / \gpratio))
  \]
  This min is maximized at $\paramratio = \critparamratio$, where the
  two terms in the min are both equal to
  \[
  \frac{\gpratio \ln \gpratio}{\gpratio - 1} + \ln (\frac{\gpratio
    \ln \gpratio}{\gpratio - 1})
  = \frac{\ln \gpratio}{\gpratio - 1} + \ln (\gpratio - 1) - \ln
  \ln \gpratio
  \]
  The desired bound follows.
\end{proofof}

\begin{proofof}{Theorem~\ref{thm:expected-error}}
  Define $\paramlo$ and $\paramhi$ as in the proof of
  Theorem~\ref{thm:worst-error}: $\paramlo =
  \frac{\paramratio}{\gpratio} \optparam, \paramhi = \paramratio
  \optparam$, for some $\paramratio \in [1, \gpratio]$.  Because
  $\log_{\gpratio} \param_0 = \unif \sim U(0,1)$, we have $\paramratio
  \sim \gpratio^{U(0,1)}$.  Apply the bound from
  Theorem~\ref{thm:error} twice, using the bound from $\paramhi$ for
  $\paramratio \leq \critparamratio$ and the bound from $\paramlo$ for
  $\paramratio \geq \critparamratio$.  Let $\unifhat = \ln_{r}
  \critparamratio$.  Then, in expectation, the better of these two
  bounds is:
  \begin{equation}
    \label{eq:exp-bound-with-integrals}
    -\logmult + \logmult \left( \int_0^{\unifhat} (\gpratio^\unif - \ln
      \gpratio^\unif) d\unif + \int_{\unifhat}^1 (\gpratio^{\unif - 1} -
      \ln \gpratio^{\unif - 1}) du \right)    
  \end{equation}
  The first integral simplifies to
  \[
  \frac{\gpratio}{\gpratio-1} - \frac{1}{\ln \gpratio} - \half
  \unifhat^2 \ln \gpratio
  \]
  and the second simplifies to
  \[
  \frac{1}{\ln \gpratio} - \frac{1}{\gpratio - 1} + (1 - \unifhat) \ln
  \gpratio - \half  (1 - \unifhat^2) \ln
  \gpratio
  \]
  Summing these gives $1 + \half \ln \gpratio - \unifhat \ln
  \gpratio$.  Plugging this back
  into~\eqref{eq:exp-bound-with-integrals} gives the 
  promised bound.
\end{proofof}

These bounds are quite tight, even for generous values of $\gpratio$.
For instance, when $\gpratio = 2$, the worst-case error from
Theorem~\ref{thm:worst-error} is only $0.5182$ dB, and our upper bound
on the expected error is a mere $0.1732$ dB.  These errors are dwarfed
by the validation errors of the model itself, as reported in Plets
\etal~\cite{DBLP:journals/ejwcn/PletsJVTM12}, which are mostly in the
1dB to 5dB range.

\subsection{Practical considerations}

We have one piece of unfinished business, which is to define the range
of $\param_i$ values for which $\GPd$ must run the $\SP(\param_i)$
computation.  In addition, we shall demonstrate some pruning tricks on
$\gTwo$ which imply, under some reasonable assumptions, that the total
running time of all of our $\SP(\param_i)$ computations is $O(1)$
times the cost of running Dijkstra just once on $\gTwo$.  Finally, we
discuss how to save time and memory by running Dijkstra on $\gTwo$
implicitly, while explicitly constructing only $\gOne$ (and not
$\gTwo$) in memory.

\subsubsection{Pruning $\gTwo$}
\label{sec:pruning}

Let us fix a particular destination $\dest$, and set
\[
\activeinterval_{\optparam} = [\frac{\critparamratio}{\gpratio}
\optparam, \critparamratio \optparam].
\]
The proofs of Theorems~\ref{thm:worst-error} and
\ref{thm:expected-error} rely only on running $\SP(\param)$ for some
$\param \in \activeinterval_{\optparam}$.  Although we do not know
$\optdist$ or $\optparam = \logmult / \optdist$, we
do know that $\optparam \in [\frac{\logmult}{\dmax}, \frac{\logmult}{\dmin}]$.
Therefore, if we run $\SP(\param_i)$ for each of the $\param_i$ in
\begin{align}
  \label{eq:activeinterval}
  \activeinterval(\dest) := \bigcup_{\optparam \in
    [\frac{\logmult}{\dmax}, \frac{\logmult}{\dmin}]} \activeinterval_{\optparam}
  = \left[\frac{\logmult \critparamratio}{\gpratio \dmax},
    \frac{\logmult \critparamratio}{\dmin} \right]
\end{align}
then we satisfy the error bounds in Theorems~\ref{thm:worst-error} and
\ref{thm:expected-error}.  Therefore, for each $\param_i$ considered
in algorithm $\GPd$, we need to include node $\dest$ in $\gTwo$ only
for the destinations
\begin{align*}
  \mpts(\param_i) &:= \{\dest : \param_i \in
  \activeinterval(\dest) \} \\&= \left \{\dest : \dmin(\dest) \leq
    \frac{\logmult \critparamratio}{\param_i} \text{ and } \dmax(\dest) \geq
    \frac{\logmult \critparamratio}{\gpratio \param_i} \right\}.  
\end{align*}
In particular, if $\mpts(\param_i) =
\emptyset$, then we do not have to run $\SP(\param_i)$ at all.  

When we first defined $\GPd$, we deferred the question of which
$\SP(\param_i)$ computations are necessary.  Let us define $\Dmin =
\min_{\dest} \dmin(\dest)$ and $\Dmax = \max_{\dest} \dmax(\dest)$.
Then we must do all $\param_i$ in $[\frac{\logmult
  \critparamratio}{\gpratio \Dmax}, \frac{\logmult
  \critparamratio}{\Dmin} ]$.

The multiplicative width of $\activeinterval(\dest)$ is only $\gpratio
\frac{\dmax}{\dmin}$, so in expectation, each destination $\dest$ is pruned from $\gTwo$ for all but $\log_{\gpratio} \frac{r
  \dmax(\dest)}{\dmin(\dest)}$ values of $\param_i$.  Recall that
$\dmin(\dest)$ is merely the straight-line distance from $\src$ to $\dest$, whereas
$\dmax(\dest)$ is the distance along the \st path $\Path_0$ with
lowest (penetration + diffraction) losses.  Since the
diffraction losses are relatively high compared to the penetration
losses, \eg for drywall, a $90^{\degree}$ turn costs the same as
penetrating 2.5 walls (Section~\ref{sec:model}).  Therefore, we would expect that path $\Path_0$
does not bend too much, and therefore
$\frac{\dmax(\dest)}{\dmin(\dest)}$ will be fairly small in practice,
typically less than 2.  In this case, if using  $\gpratio
= 2$, then we include most destinations $\dest$ in only one
or two of the $\SP(\param)$ computations.

We can also prune some of the corner points from $\gTwo$.  If the distance $\dist(\src, \corner) + \dist(\corner,
\dest)$ from $\src$ straight to corner point $\corner \in \corners$
straight to $\dest$ exceeds $\dmax(\dest)$, then we need not consider
any \st paths that go through $\corner$, because they will be both longer
than $\Path_0(\dest)$ and have equal or greater loss than
$\Path_0(\dest)$ (which has minimum loss over all \st paths).  In this
case, we may prune the edge $(\corner, \dest)$ from $\gTwo$.  In
addition, if this condition holds for all $\dest \in
\mpts(\lambda_i)$, then we prune $\corner$ from $\gTwo$ entirely.

Conceptually, our sequence of $\SP(\param_i)$ computations is
performed on a single graph, $\gTwo$.  However, the pruning operations
that we just discussed shrink it greatly for most values of
$\param_i$, because of the geometry.  Let us sweep $\param$ downward
from $\frac{\logmult \critparamratio}{\Dmin}$ down to $\frac{\logmult
  \critparamratio}{\gpratio \Dmax}$.  Initially, only the nodes close
to the source $\src$ remain unpruned.  Each time we divide $\param$
by $\gpratio$, the outer radius of the annulus of measurement points
defining $\mpts(\param)$ grows by a factor of $\gpratio$.
Assuming the ratio $\frac{\dmax(t)}{\dmin(t)}$ is $O(1)$, then the
inner radius of this annulus also grows by roughly a factor of $\gpratio$.
In the typical case, the set of measurement points is a uniform grid,
which means that $|\mpts(\param)|$ grows by roughly a factor of
$\gpratio^2$.  If the corners are also spaced relatively uniformly, then the
number of unpruned corners also grows by roughly $\gpratio^2$.
Therefore, the set of relevant unpruned $\corners-\corners$ and $\corners-\dests$
edges grows by roughly $\gpratio^4$.  Therefore, the total cost of all
of the $\SP(\param_i)$ computations is dominated by the ones at the
end of the process (small $\param_i$), where the pruned version of
$\gTwo$ is the largest, and by $\SP(0)$, which is used to compute
$\dmax(\dest)$ for all $\dest$ and so must run on the full $\gTwo$.

\subsubsection{Running Dijkstra on $\gTwo$ implicitly}
\label{sec:implicit-G2}

Recall Dijkstra's shortest path
algorithm~\cite{KleinbergTardos-algorithms-book}.  We maintain a
distance label $\dijkstralabel_v$ on each node $v$, initialized to 0
for $\src$ and $\infty$ for all other nodes, with all labels
\emph{active}.  At each step, we select the smallest active label,
make it inactive (\aka \emph{finalize} it), and \emph{relax} each of
its outgoing edges $e = (v, w)$, meaning that we update
$\dijkstralabel_w \leftarrow \min(\dijkstralabel_w, \dijkstralabel_v +
\weight_e)$.  Once the last label is finalized, $\dijkstralabel_v$ is
the weight of the shortest path from $\src$ to $v$.

Recall that most of the nodes in $\gTwo$ are sockets $(e, \corner)$
from $\gOne$, where $\corner \in \corners$ is a corner point with
incident edge $e$, and most of the edges of $\gTwo$ are the
intra-corner edges from each incoming socket to each outgoing socket
at $\corner$.  We can save a huge amount of memory by explicitly
constructing and storing only $\gOne$, and running Dijkstra implicitly
on $\gTwo$.  To do this, we maintain our distance labels
$\dijkstralabel$ on the sockets of $\gOne$.  When we finalize the
label of an incoming socket $(e, \corner)$, we must relax all of its
outgoing intra-corner edges to the outgoing sockets $(e', \corner)$.
But the cost of each of these edges is a function of just three
things: (1) the diffraction angle between $e$ and $e'$, (2) the sector
of $e$, and (3) the sector of $e'$, where \emph{sector} refers to the
directions between two consecutive walls meeting at $\corner$.  The
pair of sectors determines the penetration loss
and the deflection angle determines the diffraction loss.

This explains how to run Dijkstra without ever storing $\gTwo$.
Better yet, we can actually avoid the vast majority of no-op
relaxations (\ie ones that do not actually update their label).  This is
because, for all outgoing sockets $(e', \corner)$ within a given
sector, the penetration loss from $(e, \corner)$ is the same, and
the diffraction loss, plotted as a function of the angle $\Angle$ is a
line with constant slope equal to $\pm \diffconst_\corner$, the
diffraction constant at corner $\corner$.  Thus, we can visualize each
of the finalized incoming sockets $(e, \corner)$ at corner $\corner$
as inducing either a V-shape (for its own and its opposite sector) or
a line (for all other sectors), representing the implied path weight to
a hypothetical outgoing socket at angle $\Angle$.  The actual Dijkstra
label will be the minimum of these lines and V-shapes.

To actually perform the relaxation for a newly-finalized incoming socket
$(e, \corner)$, we start in the opposite sector at angle 0, the bottom
of the V, and march through the outgoing sockets in clockwise order to
$180^{\degree}$, then do it again counterclockwise.
If we ever encounter an angle at which our implied label exceeds the
existing label (\aka a no-op relaxation), we know that our line is
dominated for the rest of that sector, since we are increasing at rate
$\diffconst_\corner$ and all other lines are increasing or
decreasing at that same rate.  We then pick up at the next sector,
where we have a chance again because the vertical offset of each line
is different (from the differing penetration losses, depending on the
sector of the corresponding incoming socket).  
Therefore, the number of no-op Dijkstra relaxations that we must
perform is bounded by the number of sectors + 2 (since the
$0^{\degree}$ and $180^{\degree}$ sectors each count twice).



\section{Experiments}
\label{sec:experiments}

\label{sec:data}

\textbf{Datasets:}
For our experiments, we consider two types of artificial ``buildings.''  These
are not meant to replicate real buildings, but rather just to exhibit some
properties of our algorithms.  The first type are ten random ``maze'' buildings,
like the one shown in Figure~\ref{fig:heatmap}.  These are formed by taking a
20x20 grid graph, removing a random spanning tree, and then taking the planar
dual, leaving a 20x20 connected maze of 3m x 3m cells.  This gives 60m x 60m
buildings, with 441 corner points, 441 walls, and 3600 measurement points (on a
1m grid).  The second building is an artificial office building, like the one
shown in Figure~\ref{fig:paths}, to contrast with the random mazes.  Although
this is not a real office building, it does provide a check that the
experimental results are not purely an artifact of the random mazes.  It
consists of a very regular grid of 3m x 4m offices connected by 2m wide
hallways, with 12 rows of 20 offices each (where Figure~\ref{fig:paths} shows
just a portion with 6 rows of 10 offices).  This office is 62m x 60m, with 418
corner points, 658 walls, and 3720 measurement points (on a 1m grid).  For both
types of artifical buildings, the exterior walls are concrete (with 15.0dB
penetration loss), the interior walls are layered drywall (with 2.0dB
penetration loss), and the diffraction loss at all corners is 5.0dB/90\degree \
(loss values from Plets \etal~\cite{DBLP:journals/ejwcn/PletsJVTM12}).  Of
course, our algorithms work just fine with different loss parameters for each
wall and corner, but uniform values are most easily justified.

\label{sec:approx-errors}

\textbf{Approximation errors:}
First, we consider the approximation error of the GP algorithm from
Section~\ref{sec:geom-prog}.  For each of the ten random mazes, we sampled 1000
random source-destination pairs, and for the office building, we sampled 10,000
random pairs.  For each pair we computed the full convex hull of parametric
shortest paths (Section~\ref{sec:st-hull}).  Even though
Carstensen~\cite{carstensen} gives a worst-case lower bound of $n^{\Omega(\log
  n)}$ for the number of \extremepts on the convex hull, the worst case we
encountered was 19 \extremepts, and the mean was only 4.2 for the mazes, and 5.5
for the office building.

\begin{figure}
  \begin{center}
  \includegraphics*[width=0.9\linewidth,keepaspectratio=true]{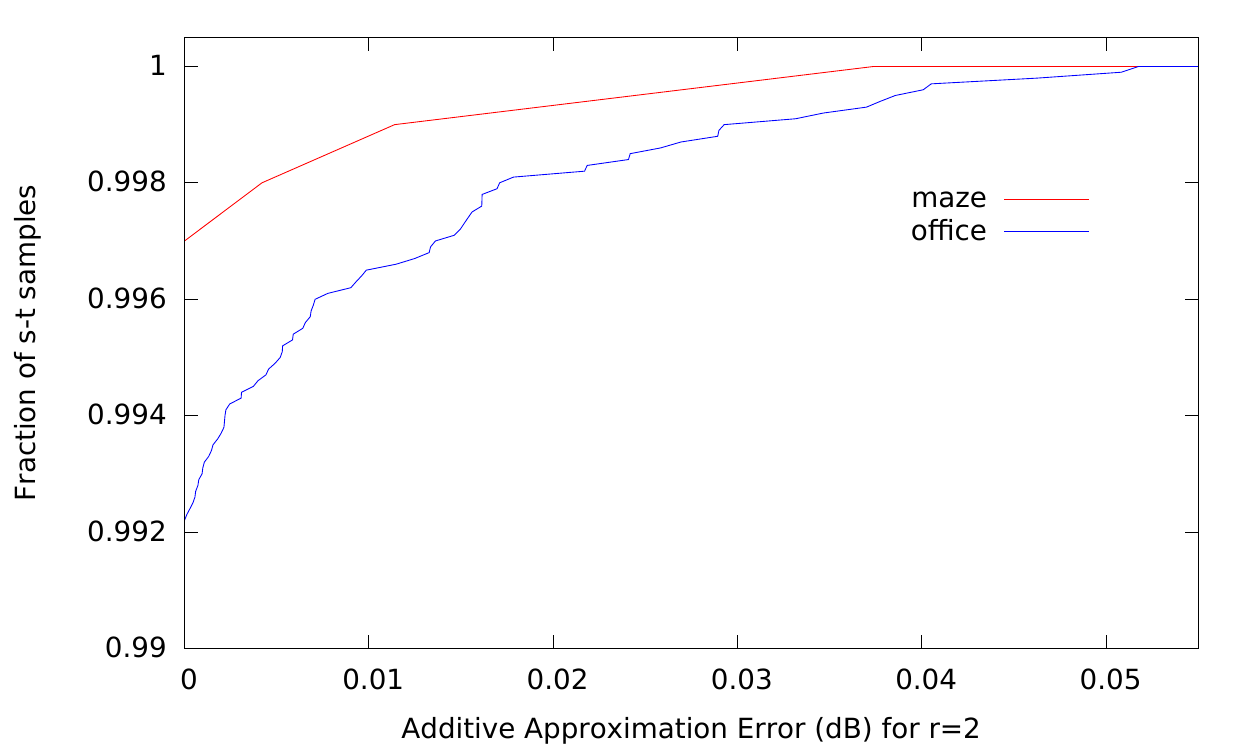}
  \includegraphics*[width=0.9\linewidth,keepaspectratio=true]{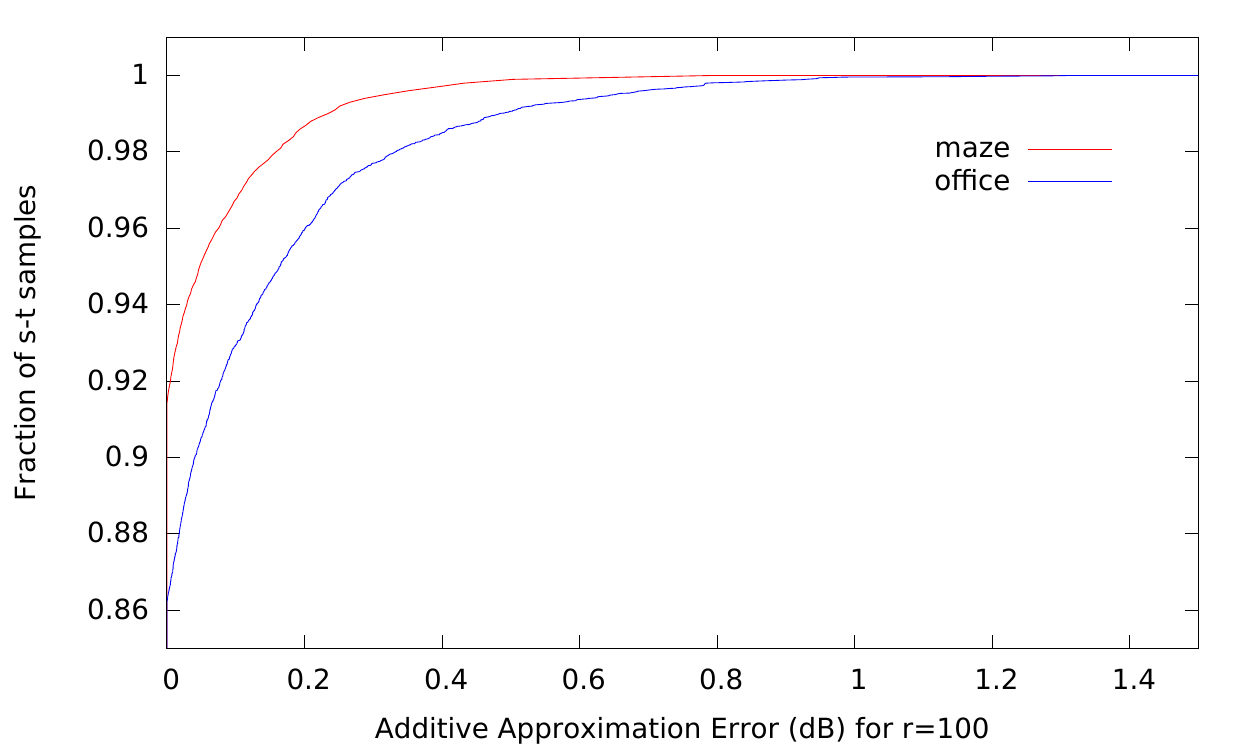}
  \caption{\label{fig:error-plots}Error distributions for $\gpratio=2$ and $\gpratio=100$}
  \end{center}
\end{figure}

\begin{figure}
	\begin{center}
		\includegraphics*[width=0.9\linewidth,keepaspectratio=true]{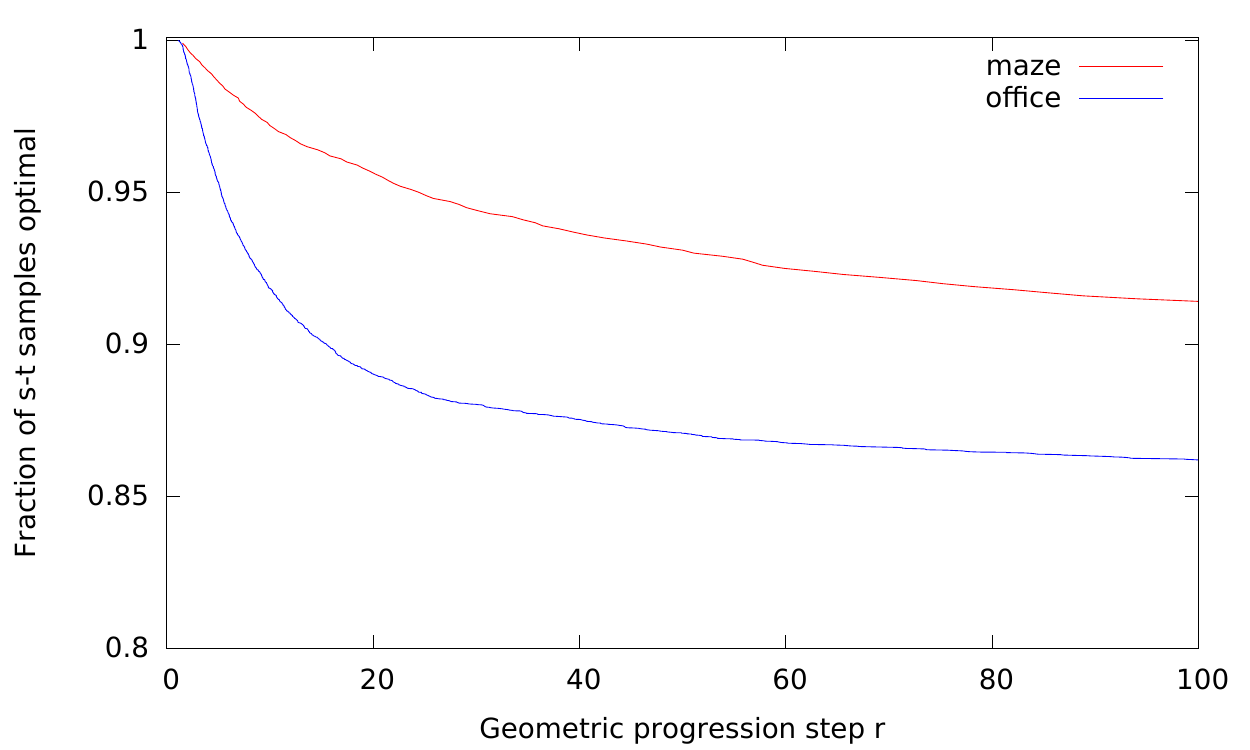}
		\caption{\label{fig:optimal-csf}Frequency of identification of dominant \st path}
	\end{center}
\end{figure}
    
Based on these convex hulls, we compute the exact solution to the IDP model,
which allows us to compute the expected error for each source/destination pair
in the GP approximation algorithm.  All expectations are \wrt the random choice
of $\param_0$ in $\GPd$.  Figure~\ref{fig:error-plots}
shows the error distributions for $\gpratio=2$ and $\gpratio=100$.  Recall that
our errors are one-sided: we can only overestimate the path loss of the dominant
path.  The actual expected errors are much better than the bounds from
Theorems~\ref{thm:worst-error} and~\ref{thm:expected-error}.  For instance, with
$\gpratio=2$, the GP algorithm failed to find the exact dominant \st path for
fewer than 0.8\% of the \st pairs, and the error exceeded 0.06dB for \emph{none}
of them.  These approximation errors are insignificant compared to the model
validation errors of 1dB to 5dB reported by Plets
\etal~\cite{DBLP:journals/ejwcn/PletsJVTM12}.  Even for the extreme case of
$\gpratio=100$, for which Theorem~\ref{thm:expected-error} gives an expected
error bound of 6.6dB, the worst observed error is only 1.5dB and 99\% of the \st
pairs have error below $0.6$dB.  There is never any reason to use such a large
value of $\gpratio$; we show it just to emphasize that our results are
extremely robust to $\gpratio$.

To understand why the geometric progression algorithm actually finds the
dominant \st path so frequently, consider the two breakpoints $\paramlo$ and
$\paramhi$ corresponding to the two segments of the convex hull adjacent to the
\extremept representing the dominant \st path
(Figure~\ref{fig:level-sets-cvx-hull}).  The dominant \st path will be returned
by $\SP(\param)$ for every value of $\param \in (\paramlo, \paramhi)$.  In
particular, if $\paramhi/\paramlo > \gpratio$, then the geometric progression is
\emph{guaranteed} to have $\param_i \in (\paramlo, \paramhi)$ for some $i$, and hence find
the dominant \st path.  Figure~\ref{fig:optimal-csf} shows, as a function of
$\gpratio$, how often we are guaranteed to find the dominant \st path, \ie what
fraction of our sampled \st pairs satisfy $\paramhi/\paramlo > \gpratio$.


\textbf{Pruning} $\mathbf{\gTwo:}$
As observed in Section~\ref{sec:pruning}, a measurement point $\dest$ needs to
be included when running $\SP(\param_i)$ only for each $\param_i \in
\activeinterval(\dest)$, so is pruned from $\gTwo$ for all but $\log_{\gpratio}
\frac{r \dmax(\dest)}{\dmin(\dest)}$ values of $\param_i$ (in expectation).  To
evaluate how effective this pruning was, we considered $\gpratio=2$ for ten
random choices of $\src$ and a 1m grid of measurement points for each of ten
random maze buildings and the office building.  For the random maze buildings,
the expected number of $\SP(\param_i)$ computations that left the average
measurement point unpruned was only 1.06, and the maximum expectation we
encountered over all measurement points was only 2.14.  For the office building,
the long straight hallways result in higher $\frac{\dmax(\dest)}{\dmin(\dest)}$
ratios, but still the average number of $\SP(\param_i)$ a measurement point was
included in was only 1.29, and the maximum we encountered was only 2.69.  Hence,
the total complexity of the full $\GP{2}{\param_0}$ algorithm is very close to
the complexity of running two Dijkstra computations on the unpruned version of
$gTwo$: one with $\param = 0$ to compute $\dmax(\dest)$ for all $\dest$, and the
sequence of Dijkstra runs on pruned versions of $gTwo$ add up to about one
additional Dijkstra on the full $\gTwo$.


\textbf{Implicit $\gTwo$ savings:}
A key implementation detail is to avoid no-op Dijkstra edge relaxations on the
implicit representation of $\gTwo$ (Section~\ref{sec:implicit-G2}).  In the run
that generated the heat map in Figure~\ref{fig:heatmap}, 
roughly 99\% of the relaxations were no-ops, so this trick allowed us to perform
only $2.18\times 10^8$ relaxation steps rather than $2.87\times 10^{10}$.

\textbf{Running time:}
For the mazes, building $\gTwo$ took 3.0 CPU sec, and generating a
heat map from a single source took 1.0sec.  For the office, building
$\gTwo$ took 2.9sec, and a single heat map took 1.3sec.  The experiments were run
using a single thread on a 3.6GHz Intel Xeon E5-1650v4 CPU.



\section{Smoothed Analysis}
\label{sec:exact-alg}

\newcommand{\flowe}{x_e}
\newcommand{\unitv}{u}
\newcommand{\spannedspace}{V}
\newcommand{\spannedspaceU}{U}
\newcommand{\dimn}{n}

The exact algorithm runs fast in simulations, despite its worst-case super
polynomial running time in theory.  The good practical performance can be
explained theoretically by smoothed analysis: the idea that the worst-case
instances are rare and most likely in practice the algorithm encounters ``good"
instances, for which the number of breakpoints (and hence also the number of
\extremepts) that are enumerated by the exact algorithm, is small.  We state
this result informally, then follow up with a rigorous development.

\begin{thm}\label{thm:smooth}
  The exact IDP model can be solved in smoothed polynomial time.
\end{thm}

Formally, the smoothed analysis framework defines a perturbation of the input
(the edge weight vectors $\dist, \loss \in \Rm$, where $\dimm$ is the number of
edges in the graph) and proves that in expectation with respect to this
perturbation, the algorithm running time (specifically, the number of
\extremepts), is polynomial.

To help us state the smoothed bound, we first note that the optimal path is an \extremept of the flow polytope $\Path \in \Rm$, defined via the standard flow constraints: 
$$
\begin{array}{ll}
\displaystyle{ \sum_{e \ \text{ out of } \ v}{\flowe} = \sum_{e \ \text{ into } \ v}{\flowe} }& 
      \textrm{ for each node } v \neq \src,\dest \\
\displaystyle{ \sum_{e \ \text{ out of } \ \src}{\flowe} = 1 } & \textrm{ for source } \src \\
\displaystyle{ \sum_{e \ \text{ into } \ \dest}{\flowe} = 1 } 
       & \textrm{ for destination } \dest 
\vspace{2mm}
\\ 
\displaystyle{ 0 \leq x_e \leq 1 } & \textrm{ for each edge }e,
\end{array}
$$
where $x_e$ is the flow on edge $e$.

The exact algorithm projects the polytope $\Path$ onto the 2-dimensional plane spanned by the distance and loss vectors $\dist,\loss$ and enumerates \extremepts of the 2D-polytope projection. The smoothed bound below implies that when the distances and losses $\dist,\loss$ are perturbed slightly, then the number of \extremepts\ on the 2D-polytope projection is polynomial in expectation with respect to the perturbation.   We use the following perturbation and theorem from~\cite{nikolova-ssp} that applies directly to our problem.
  
\begin{defn}[Perturbation \cite{nikolova-ssp}]
For any $\rho>0$ and any unit vector $\unitv$, we define a \emph{$\rho$-perturbation
of $\unitv$} to be a random unit vector $v$ chosen as follows:\\
$1)$ Randomly select an angle $\theta\in[0,\pi]$ from the exponential
distribution of expectation $\rho$, restricted to the range $[0,\pi]$.
\\
$2)$ Choose $v$ uniformly at random from the set of vectors that make an angle
of $\theta$ with $\unitv$.
\end{defn}

The smoothed bound in the theorem below states that the expected number of extreme points of the projection of the feasible polytope $P$ onto the perturbed plane spanned by vectors $v_1, v_2$ is polynomial:
The smoothed bound on the number of \extremepts is given in the following theorem: 

\begin{thm}[Smoothed bound~\cite{nikolova-ssp}]\label{smoothed}
Let $\unitv_1$ and $\unitv_2$ be arbitrary unit vectors and denote $\spannedspaceU=\Span{\unitv_1,\unitv_2}$.
Let $\spannedspaceU=\Span{\unitv_1,\unitv_2}$ be an arbitrary 2-plane.  
Let $v_1$ and $v_2$ be $\rho$-perturbations of $\unitv_1$ and $\unitv_2$, respectively, and let
$\spannedspace=\Span{v_1,v_2}$.  The expected number of edges of the projection of $\Path$ onto $\spannedspace$ is at most $4\pi\sqrt{2 \dimm}/\rho$, for $\rho < 1/\sqrt{\dimm}$.
\end{thm}

There is a tradeoff between the size of the perturbation and the bound in the theorem---the smaller the perturbation, the larger the bound on the expected number of \extremepts.  However, if we set 
$\rho = \frac{1}{\sqrt{2\dimm}}$, for example, we get the bound $8\pi \dimm$, which is linear in the number of edges $\dimm$.  
By the one-to-one correspondence between \extremepts on the 2D-polytope projection and breakpoints in the parametric shortest paths problem,  
the smoothed bound above implies the expected polynomial complexity of the exact algorithm, as stated in the following corollary: 

\begin{cor}
If the edge weight vectors $\dist, \loss \in \Rm$ are fixed vectors and $\dist',\loss' \in \Rm$ denote their $\rho$-perturbations, then the expected number of \extremepts on the path polytope shadow on $\Span{\dist',\loss'}$ is linear in the number of graph edges $\dimm$ and consequently the $\dimn^{\Theta(\log \dimn)}$ exact algorithm for the dominant path problem has expected polynomial running time. 
\end{cor}

This Corollary is the formal statement of Theorem~\ref{smoothed}.



\section{Future work}
\label{sec:future-work}

This paper focuses on algorithm design, not algorithm engineering.  Although our
prototype implementation is reasonable, it has not been highly engineered for
speed.  We could accelerate it by pruning edges of $\gOne$ above some loss
threshold, handling measurement points outside the main Dijkstra loop and
priority queue, tuning data structures, cache optimization, etc.  After such
improvements, a careful "horserace" running time comparison against tree-search
dominant path codes might be appropriate.  The current paper instead focuses on
proving the GP algorithm's viability, owing to its already-fast running time and superb
fidelity to the exact \idp model.

For simplicity, we focused on the 2D indoor dominant path model, but the outdoor
and mixed models are also important.  It would be interesting to apply the GP
algorithm to these models, and also to 3D models.  Finally, we hope that our
methods will be integrated into wireless nework planning tools, to support AP
placement optimization as described in Section~\ref{sec:intro}.



\section{Acknowledgments}

This paper is dedicated to our late co-author David S. Johnson, an exemplary
friend, mentor, and human being.  This work was partially supported by NSF through grants CCF-1216103, CCF-1331863, CCF-1350823 and CCF-1733832.
Mikkel Thorup's research is supported by his Advanced Grant DFF-0602-02499B from the Danish Council for Independent Research and by his Investigator Grant 16582, Basic Algorithms Research Copenhagen (BARC), from the VILLUM Foundation.



\bibliographystyle{ACM-Reference-Format}
\bibliography{dompath} 


\end{document}